\documentclass[submission,copyright,creativecommons]{eptcs}
\providecommand{\event}{PLACES 2026} 

\usepackage[T1]{fontenc}

\usepackage{ifthen}
\usepackage{multicol}
\usepackage{pifont}
\usepackage{stmaryrd}
\usepackage{latexsym}
\usepackage{amsmath,amssymb,amsthm}
\usepackage{amsfonts}
\usepackage{alltt}
\usepackage{tabularx}
\usepackage{proof}
\usepackage{xfrac}
\usepackage{comment}
\usepackage{xcolor}
\usepackage{DotArrow}
\usepackage{xspace}
\usepackage{ulem}
\usepackage{txfonts}
\usepackage{hyperref}

\usepackage[scaled=0.87]{helvet}

\newtheorem{definition}{Definition}

\newtheorem{theorem}{Theorem}

\newtheorem{lemma}{Lemma}
\newcounter{remark}

\title{Determinacy with Priorities up to Clocks}

\author{
Luigi Liquori
\institute{Centre Inria de l'Université Côte d'Azur}
\and
Michael Mendler
\institute{University of Bamberg}
\and
Claude Stolze
\institute{University of Bamberg}
}

\newcommand{\titlerunning}{Determinacy with Priorities up to Clocks}
\newcommand{\authorrunning}{L. Liquori, M. Mendler, C. Stolze}

\hypersetup{
  bookmarksnumbered,
  pdftitle    = {\titlerunning},
  pdfauthor   = {\authorrunning},
  pdfkeywords = {Timed process algebra, Synchronous programming}
}

\newcommand{\rmm}[1]{}

\newcommand{\eqdef}{\stackrel{_\mathit{def}}{=}}

\newcommand{\bip}{\textsf{BIP}\xspace}
\newcommand{\reo}{\textsf{REO}\xspace}

\newcommand{\ccs}{\textsf{CCS}\xspace}
\newcommand{\csp}{\textsf{CSP}\xspace}
\newcommand{\csa}{\textsf{CSA}\xspace}
\newcommand{\pmc}{\textsf{PMC}\xspace}
\newcommand{\ccslm}{\textsf{CCS}$^{{\sf spt}}$\xspace}
\newcommand{\ccsp}{\textsf{CPG}\xspace}

\newcommand{\cP}{\ensuremath{\mathcal{P}}\xspace}
\newcommand{\cH}{\ensuremath{\mathcal{H}}\xspace}

\newcommand{\tcode}[1]{{\mathsf{#1}}}

\newcommand{\ol}[1]{\ensuremath{\overline{#1}}}
\newcommand{\col}{\ensuremath{{:}}}
\newcommand{\cpar}{\,|\,}
\newcommand{\cseq}{\ensuremath{\tcode{.}}\xspace}

\newcommand{\zero}{{\tcode{0}}\xspace}
\newcommand{\clockzero}{{\eset}}
\newcommand{\clockone}{{\{\sigma\}}}

\newcommand{\tsf}[1]{\textsf{#1}}

\newcommand{\A}{\mathcal{A}}
\newcommand{\I}{\mathcal{I}}
\newcommand{\R}{\mathcal{R}}

\renewcommand{\L}{\mathcal{L}}
\newcommand{\C}{\mathcal{C}}

\newcommand{\Act}{\mathcal{L}^{\tau}}
\newcommand{\Thread}{\mathcal{S}}
\newcommand{\Term}{\mathcal{T}}
\newcommand{\Proc}{\mathcal{P}}
\renewcommand{\P}{\mathcal{P}}

\newcommand{\coA}{\overline{\mathcal{A}}}

\newcommand{\wilA}[1]{\ensuremath{\tsf{iA}^{#1}}\xspace}
\newcommand{\iAstar}{\ensuremath{\textsf{iA}^{\ast}}\xspace}
\newcommand{\clock}{\ensuremath{\textsf{clock}}}
\newcommand{\iA}{\ensuremath{\textsf{iA}}\xspace}
\newcommand{\eset}{{\{\}}}

\newcommand{\Derives}[3]
  {\mbox{ \ensuremath{\xrightarrow[{\protect\raisebox{-0.25em}[0em][0em]{$^{#3}$}}]
  {\protect\raisebox{-0.5em}[0em][0em]{$^{#1}$}}\!\!{_{#2}}\, }}}

\newcommand{\fstep}[1]
  {\ensuremath{\mathrel{\stackrel{#1}{\rightarrow}_{^{\xspace}}}}}

\newcommand{\ActR}{\ensuremath{\textit{Act}}\xspace}
\newcommand{\RestrR}{\ensuremath{\textit{Restr}}\xspace}
\newcommand{\SumR}{\ensuremath{\textit{Sum}}\xspace}
\newcommand{\ParR}{\ensuremath{\textit{Par}}\xspace}
\newcommand{\ConR}{\ensuremath{\textit{Con}}\xspace}
\newcommand{\ComR}{\ensuremath{\textit{Com}}\xspace}

\newcommand{\restrict}{\ensuremath{\,\backslash\,}}
\newcommand{\ABRO}{\textsf{ABRO}\xspace}
\newcommand{\CR}{\textsf{DP}\xspace}
\newcommand{\eschews}{\ensuremath{\textsf{eschews}}}

\begin{document}
\maketitle


\begin{abstract}
  In Milner's seminal book on communication and concurrency introducing \ccs, a process algebra inherently non-deterministic, chapter 11 was completely devoted to introduce the notion of determinacy and confluence in order to identify a subcalculus of \ccs in which all definable agents are confluent.
  At the same time, or shortly later, determinate semantics were given for programming languages that reconcile concurrency and determinacy, such as \textsf{Esterel} by Berry and Gonthier, or \textsf{SL} by Boussinot and de Simone. These dedicated semantics do not easily map to Milner's confluence theory for \ccs, which is unable to express causality and shared memory multi-threading with reaction to absence in a compositional way.
  We present an extension of \ccs with priority-guarded actions and clocks, and we exploit the added expressiveness to enrich Milner's original notion of confluence by the new concept of \textit{coherence} which permits us to encode, in a compositional fashion, synchronous programming languages such as Esterel.
 \end{abstract}



\section{Introduction}\label{sec:intro}
 According to Milner~\cite{Milner:CCS}, the notion of determinacy is tied up with predictability: {\it ``if we perform the same experiment twice on a determinate system -- starting each time in its initial state -- then we expect to get the same result, or behaviour, each time.''}
In $\lambda$-calculus, determinacy refers to the uniqueness  of normal forms which is a result of the Church-Rosser Property. In concurrency, it corresponds to the absence of race-conditions, which are the source of many bugs in concurrent programming. In this paper we adopt the setting and notation of Milner's \ccs, where we have action-labelled transitions $P \Derives{\alpha}{}{} Q$ (which is either the strong or weak transition), compatible with a suitable congruence $\cong$  that preserves normal forms.
Milner defines determinacy~\cite{Milner:CCS} (Def.~2 and Def.~3, Chap.~11.1) as the condition that any given action can only lead to congruent continuations.

\begin{definition}
 A process $P$ is \textit{determinate} (modulo $\cong$) if for all its derivatives $Q$ and action $\alpha \in \Act$, if $Q \Derives{\alpha}{}{} Q_1$ and $Q \Derives{\alpha}{}{} Q_2$ then $Q_1 \cong Q_2$.
\label{def:determinacy}
\end{definition}

The problem with determinacy, defined in this way, is that it is not closed under parallel composition. The solution proposed by Milner~\cite{Milner:CCS} (Chap.~11.3), is to strengthen determinacy to the notion of \textit{confluence} which turns out to be closed for parallel composition under natural restrictions.
\begin{definition}
  $P$ is \textit{confluent} (modulo $\cong$) if it is determinate and for every derivative $Q$ of $P$ with transitions $Q \Derives{\alpha_1}{}{} Q_1$ and $Q \Derives{\alpha_2}{}{} Q_2$ such that $\alpha_1 \neq \alpha_2$, there exist $Q_1' \cong Q_2'$ such that $Q_1 \Derives{\alpha_2}{}{} Q_1'$ and $Q_2 \Derives{\alpha_1}{}{} Q_2'$.
\label{def:local-confluence}
\end{definition}

\noindent Definition~\ref{def:local-confluence} subsumes \textit{strong confluence}~\cite{Milner:CCS} (Def.~4, Chap.~11.3) when the transitions are strong and $\cong$ is a bisimulation ($\sim$), and \textit{weak confluence}~\cite{Milner:CCS} (Def.~5, Chap.~11.3) for weak transitions and observation equivalence ($\approx$). For instance, consider the process $P = (R_1 \cpar S \cpar R_2) \restrict r$ with components $R_1 = \ol{r} \cseq a \cseq \zero$, $S = r \cseq S$, and $R_2 = \ol{r} \cseq b \cseq \zero$. Process $P$ is not strongly confluent since it is not determinate (i.e., modulo $\sim$):
\begin{displaymath}
  P \Derives{\tau}{}{} (a \cseq \zero \cpar S\cpar \ol{r} \cseq b \cseq \zero) \restrict r
  \text{ and }
  P \Derives{\tau}{}{} (\ol{r} \cseq a \cseq \zero \cpar S \cpar b \cseq \zero) \restrict r
  \text{ and }
  (a \cseq \zero \cpar S \cpar \ol{r} \cseq b \cseq \zero) \restrict r  \not\sim (\ol{r} \cseq a \cseq \zero \cpar S \cpar b \cseq \zero) \restrict r.
\end{displaymath}
However, all components $R_1$, $S$, and $R_2$ of $P$ are strongly confluent. Hence, like determinacy, strong confluence is not closed under parallel composition. On the other hand, $P$ is weakly confluent (i.e., modulo $\approx$) and converges weakly to the normal form $(a \cseq \zero \cpar S \cpar b \cseq \zero) \restrict r$.
Milner~\cite{Milner:CCS} (Prop.~17, Chap 11.4) shows that weak confluence is preserved by
\textit{confluent composition} $P \mathrel{{\cpar}_{\! L}} Q \eqdef (P \cpar Q) \restrict L,$
combining the parallel and restriction operator, where $L$ is a set of restricted labels, subject to the \textit{Separation Condition} that $\L(P) \cap \L(Q) = \eset \text{ and } \ol{\L(P)} \cap \L(Q) = L \cup \ol{L}$, where $\L(\_)$ is the function calculating the sort of a process. The former condition says that $P$ and $Q$ operate on disjoint actions and the latter says that every possible communication between $P$ and $Q$ is restricted by $L$. This is a form of sort separation, ensuring that every action has at most one synchronisation partner either inside or outside of $(P \cpar Q) \restrict L$.

It may be the case that many practical examples of determinate systems can be understood as sort-separated compositions of confluent processes, for the right choice of coding scheme.
For application to synchronous programming, however, stronger compositionality arguments are needed.
For instance, replace the process $S$ in our example by
$S = r \cseq S + w \cseq \zero$ which models a simple write-once memory with read and write actions $r$ and $w$, respectively. It permits multiple readers to synchronise on $r$, not changing its state, and a single writer on $w$, leading to termination.
The process $P = (R_1 \cpar S \cpar R_2) \restrict \{r,w\}$ is weakly confluent but falls outside of Milner's confluence class:
Firstly, it violates the Separation Condition and secondly the subprocess $S$ is not confluent. Milner's result~\cite{Milner:CCS} (Prop.~17, Chap. 11.4) does not help us to verify that shared memory multi-threading without data races, like $P$, is confluent.
Memory processes that permit destructive update
are firstly not confluent and, secondly, Milner's confluent composition forbids direct multi-cast communication, because labels such as $r$ in $S$ could not be shared by two readers $R_1$, $R_2$ due to sort-separation. This means that concurrent programming languages that support shared memory  and yet have determinate reduction semantics, cannot be handled.

In a working paper~\cite{fscd}, we define \ccslm, a process algebra extending Milner's \ccs with clocks and priorities on actions up-to clocks. This process algebra is able to capture multi-clock synchronous processes in a compositional way, defining the scheduling of the processes through priorities. As any process algebra, \ccslm is nondeterministic, but deterministic programs, like e.g., Esterel's \ABRO~\cite{Berry99} can be expressed in a compositionally elegant way.
Starting from a fragment of \ccslm with priority-guarded actions and a single clock, we define a new notion, \textit{coherence}, that addresses the issues with the classical notions of determinacy and confluence mentioned above. It
has the quite nice property
that deterministic shared memory and reaction to absence, as used in synchronous programming, can be modelled.  In short: \ccslm coherent expressions can set up a minimal, telescopic foundation for semantic of synchronous programming languages. We only consider a single clock in this paper for illustration and to set the scene for application in synchronous programming. Yet, even in the absence of clocks, the notion of coherence is a non-trivial extension of Milner's confluence theory.
For lack of space, all proofs are omitted and will appear in the full version of the paper.



\section{The Syntax and Semantics of Single-clock \ccslm}
\label{sec:priorities}

We assume the reader is familiar with the notation of \textit{synchronous process algebras}~\cite{SCCS}.
We first put in place the syntactic signature of action labels and then discuss our new enriched concept of \textit{strategic} transitions that add priorities and clocks to \ccs for scheduling.

Let $\coA/\A$ be sets of \textit{co/channel names} with $a,b, \ldots$ ranging over $\A$. We will refer to the $a \in \R \eqdef \A \cup \coA$ as \textit{rendezvous actions}. Let $\C = \{\sigma\}$ be a singleton set of broadcast \textit{clock names},
disjoint from $\R$. For the sake of simplicity, this paper deals with a single clock $\sigma$.  We have $a \cpar \ol{a} = \tau$, $\sigma \cpar \sigma = \sigma$ and $\ol{\sigma} = \sigma$. Let $\L = \R \cup \C$ be the set of \textit{labels} and let $\ell$ range over $\L$, while $L,H$ range over subsets of $\L$. We write $\ol{L}$ for the set $\{\ol{\ell} \mid \ell \in L\}$. Let $\Act = \L \cup \{\tau\}$ be the set of all \textit{actions} obtained by adjoining the \textit{silent action} $\tau \notin \L$ and let $\alpha$ range over $\Act$.
All symbols can appear indexed. \ccslm terms $\Term = \Proc \cup \Thread$ come in two mutually recursive syntactic forms, \textit{processes} $\Proc$ and \textit{threads} $\Thread$.
Let $\I \subset \P$ be a set of process names, and let $\mathsf{p}, \mathsf{q} \ldots$ range over $\I$. The process terms $P, Q \in \Proc$ and the threads terms $M, N \in \Thread$ are defined by the following abstract syntax:
\begin{displaymath}
 \begin{array}[t]{lcll@{\quad}llcll@{\quad}l}
 P,Q &::=&
 \mathsf{p}
 & & \text{process name}\\
 &\mid& P \cpar Q
 & & \text{parallel composition} \\
 &\mid& P \restrict A
 & & \text{restriction}, A \subseteq \A. \\
 &\mid& M
 & & \text{thread}.
\end{array}
 \
 \begin{array}[t]{lcll@{\quad}llcll@{\quad}l}
 M, N & ::= & \zero_C
 & & \text{inactive process} \\
 &\mid& \alpha\col L \cseq P
 & & \text{prefix}, \alpha \in \Act, L \subseteq \L \\
 &\mid& M + N
 & & \text{sum}
 \end{array}
\end{displaymath}
Each restriction $P \restrict A$ acts as a name-binder where the channels $\A$ are locally bound and thus no longer free. Each $P \in \cP$ has an associated set of \textit{free labels} $\L(P) \subseteq \L$ and a \textit{clock horizon}, or simply a \textit{clock}, $\clock(P) \in \cH$ where $\cH \eqdef \{ \clockzero, \clockone \}$. A process of horizon $\clockone \in \cH$ has to synchronise on every tick of the clock. A process of horizon $\clockzero \in \cH$ does not participate in the clock and so runs asynchronously. As usual, unary operators take precedence over binary operators. The clock of a process $P$, is defined as $\clock(P) = \L(P) \cap \C$. Let us inspect the threads:
\begin{itemize}
\item $\zero_C$ is the inactive process indexed with its horizon $C$. The inactive $\zero_{\clockzero}$ has horizon $\clockzero$ and thus is asynchronous. The inactive process $\zero_{\clockone}$ with horizon $\clockone$ is synchronous and prevents other the synchronous processes from performing $\sigma$;
\item $\alpha\col L \cseq P$ is the process that may become $P$ after performing the action $\alpha$, and the set $L \subseteq \L$ contains all the actions taking precedence over it. An action $\alpha\in\R$ denotes a \ccs-style rendezvous (or handshake) action. An action $\alpha \in \C$ denotes a \csp-style broadcast (or clock) action, that shall synchronise with all the surrounding processes in the scope where this clock has been declared;
Where $L = \emptyset$, we often simply write the prefix as $a \cseq P$ instead of $a \col \eset \cseq P$;

\item $M + N$ is, as usual, the sum of $M$ and $N$, that is, it progresses either as $M$ or $N$.

\end{itemize}
Now, let us inspect the processes:
\begin{itemize}

\item $\mathsf{p}$ is name that refers to a predefined process. Names are used for recursion, e.g., $\mathsf{p} \eqdef a \col L \cseq \mathsf{p}$ is a process that infinitely offers $a$ with precedences $L$;

\item $P \cpar Q$ is, as usual,  the parallel composition of $P$ and $Q$;

\item $P \restrict A$ denotes action restriction; it makes local the rendezvous action in $A$.

\end{itemize}
A process $P$ has to synchronise on every tick of clocks in $\clock(P)$, and only those clocks: this corresponds to what Hoare~\cite{Hoare:CSP} calls the alphabet of a \csp process. It can be easily seen that direct subprocesses of a thread should have the same clock horizon as the thread, i.e. $\clock(M + N) = \clock(M) = \clock(N)$, and $\clock(\alpha\col L\cseq P) = \clock(P)$. Note also that a process $P$ of horizon $\clockzero$ cannot perform $\sigma$. It can be also noted that $\clock(P\cpar Q) = \clock(P) \cup \clock(Q)$, $\clock(P\restrict A) = \clock(P)$, and $\clock(\mathsf{p}) = \clock(P)$ whenever $\mathsf{p} \eqdef P$.

\subsection{Single-clock \ccslm: Transitions under Blocking and Prediction}

\noindent We add priority-based scheduling constraints to the standard transitions $P \Derives{\alpha}{}{} Q$ of \ccs to generate \textit{strategic transitions} $$P \Derives{\alpha}{\iota}{B} Q$$ between processes $P, Q \in \cP$, expressing that ``$P$ performs the action $\alpha$ with blocking $B$ and prediction $\iota$, and becomes $Q$''.  Formally, $\alpha \in \Act$ is the standard action of the transition, $B \subseteq \cH \times 2^\L$ is the \textit{blocking relation}, and $\iota \in \cH \rightarrow 2^\L$ the \textit{prediction function}.  For compactness, we use the notation $\alpha \col B[\iota]$ for the label of a strategic transition, referred to as a \textit{strategic label}.

\begin{itemize}

 \item Each element $(C, L) \in B$ is a blocking constraint from a thread participating in the transition $\alpha \col B[\iota]$, that only participates provided there cannot be a synchronisation on any of the actions $L$, within the clock horizon $C$. The horizon $\clockone$ consists of all actions possibly taken before the clock $\sigma$ is executed, i.e., it does not include any action happening after the first occurrence of $\sigma$. In contrast, the horizon $\clockzero$ consists of all actions both before \textit{or after} $\sigma$.

 \item The prediction $\iota$ takes a clock horizon $C$ and returns the set of action labels $\iota(C) \subseteq \L$ inside $P$ that are considered in competition to $\alpha\col B[\iota]$ within the horizon $C$. For instance, in $(a\cseq e + b\cseq f) \cpar c\cseq g$, the action $a$ is in competition with $b$, $c$ and $g$, but neither $e$ nor $f$.

\item Blocking and prediction meet each other in the scheduling of a parallel composition: The synchronisation of a transition $\alpha\col B_1[\iota_1]$ with a concurrent partner transition $\ol{\alpha}\col B_2[\iota_2]$ is unblocked if $\iota_1$ eschews $B_2$ and $\iota_2$ eschews $B_1$. More precisely, $\iota$ \textit{eschews}\footnote{This term is taken from Phillips~\cite{Phillips08}, where it plays essentially the same role. While in~\cite{Phillips08} the prediction (called ``offerings'') and blocking are simply sets of actions, in our synchronous setting they are also scoped by clock horizons.} $B$ iff for all $(C,L) \in B$ we have disjointness $\iota(C) \cap \ol{L} = \eset$, expressing that the prediction $\iota$ has no labels synchronising with $L$ in horizon $C$. We note this property $\eschews(\iota,B)$.
\end{itemize}

\noindent
Each blocking $B$ induces a function $[B] \in \cH \to 2^\L$ collecting the blocking actions within horizon $C$, defined $[B](C) \eqdef \{ \ell \mid \exists (C', L) \in B,\, C' \subseteq C, \ell \in L \}$. This relation is monotonic, which means that blocking constraints are preserved when we add clocks.
From this we obtain a natural partial ordering $B \sqsubseteq B'$ between blocking relations, as the inclusion $[B](C) \subseteq [B'](C)$ for all $C \in \cH$.
This expresses that $B$ is less restrictive, i.e, $B$ does not block in more contexts than $B'$.
The corresponding ordering on predictions is simply point-wise inclusion: $\iota_1 \sqsubseteq \iota_2$ iff $\iota_1(C) \subseteq \iota_2(C)$ for all $C \in \cH$. Observe that $\eschews$ is antitonic in its arguments, i.e., if $\iota' \sqsubseteq \iota$ and $B' \sqsubseteq B$, then $\eschews(\iota, B)$ implies $\eschews(\iota', B')$.

\begin{figure}[t]
\begin{displaymath}
  \begin{array}[b]{l}
    \infer[(\ConR)]
	  {\mathsf{p} \Derives{\alpha}{\iota}{B} Q}
 	  {\mathsf{p} \eqdef P & P \Derives{\alpha}{\iota}{B} Q} \quad
   \infer[(\ComR)]{P_1 \cpar P_2 \Derives{\ell \cpar \ol{\ell}}{\iota_{1} + \iota_{2}} {B_{1} \cup B_{2}} P_1' \cpar P_2'}{
     \begin{array}{l}
       P_1 \Derives{\ell}{\iota_{1}}{B_{1}} P_1' \quad P_2 \Derives{\ol{\ell}}{\iota_{2}}{B_{2}} P_2' \\
       \forall i\in\{1,2\},\ \eschews(\iota_{i},B_{3-i})
     \end{array}
   }
   \quad
   \infer[(\ParR)]
 	 {P \cpar Q \Derives{\alpha}{\iota + \iAstar_{-}(Q)}{B} P' \cpar Q}
	 {\begin{array}{l} \eschews(\iAstar_{-}(Q), B)\\
	     P \Derives{\alpha}{\iota}{B} P' \quad \alpha \not\in \clock(Q) \end{array}}
         \\[5mm]
         \infer[(\ActR)]{\alpha\col L\cseq P \Derives{\alpha}{\iota^{\eset}}{\{(\clock(P),L)\}} P}{}
         \quad
         \infer[(\SumR)]{M + N \Derives{\alpha}{\iota \cup (\iA(N) {-} \{\alpha\})}{B} P}{M \Derives{\alpha}{\iota}{B} P}
         \quad
         \infer[(\RestrR)]
 	       {P \restrict A \Derives{\alpha}{\iota \restrict A}{B \restrict A} Q \restrict A}
	       {P \Derives{\alpha}{\iota}{B} Q & \alpha \not\in A \cup \ol{A}}
 \end{array}
\end{displaymath}
\caption{Labelled Transition System (LTS) for \ccslm.}
\label{fig:compatible-rules}
\end{figure}

\subsection{Labelled Transition System}\label{sec:LTS}

Fig.~\ref{fig:compatible-rules} presents the labelled transition system (LTS) which formally has type $\cP \times \Act \times (2^{\cH \times 2^\L}) \times (\cH \rightarrow 2^\L) \times \cP$. A web artefact \footnote{\url{https://cstolze.github.io/synpasite/}} is available: it implements a parser and the LTS for \ccslm.
We work under the following constructivity assumptions:

\smallskip 

\noindent
\textbf{Prediction function:} $\iAstar_{C}(P)$ is the set of all the labels which appear syntactically in $P$, up-to some clock in $C$. For instance, $\iAstar_{\{\sigma\}}(a\cseq b \cseq \sigma \cseq c\cpar \ol{a}) = \{a,b,\sigma,\ol{a}\}$. Observe that $c$, which only occurs after $\sigma$, is not included.
The presence of process names makes the definition a bit difficult, so we first define the prediction function $\wilA{n}_C(P) \subseteq \L$ with an upper bound $n \in \mathbb{N}$ on the number of unfoldings of process names:
\begin{displaymath}
 \begin{array}{rcl}
 \wilA{n}_C(\zero_{C'}) & = & \eset \\
 \wilA{n}_C(\alpha \col L.P) & = &
 	\left\{ \begin{array}{ll} \{\alpha\} & \mbox{ if } \alpha \in C\\
 	\{\alpha\} \cup \wilA{n}_C(P) & \mbox{ otherwise} \end{array}\right.
	\\
 \wilA{n}_C(P\restrict A) & = & \wilA{n}_C(P) {-} (A \cup \ol{A}) \\
 \end{array}
 \qquad
 \begin{array}{rcl}
 \wilA{0}_C(\mathsf{p}) & = & \eset \\[1.5mm]
 \wilA{n+1}_C(\mathsf{p}) & = & \wilA{n}_C(P) \quad \mbox{ if } \mathsf{p} \eqdef P \\[1.5mm]
 \wilA{n}_C(P \cpar Q) & = & \wilA{n}_C(P) \cup \wilA{n}_C(Q) \\[1.5mm]
 \wilA{n}_C(M + N) & = & \wilA{n}_C(M) \cup \wilA{n}_C(N)
 \end{array}
\end{displaymath}
We define $\iAstar_C(P) = \bigcup_{n \in \mathbb{N}} \wilA{n}_C(P)$, noting that $\wilA{n}_C(P)$ is monotonic for $n$, i.e. $\wilA{n}_C(P) \subseteq \wilA{n+1}_C(P)$.

\smallskip 

\noindent \textbf{Initial actions:} $\iA(M) \subseteq \Act$ is the set of all the labels which appear syntactically at the beginning of a thread $M$. For instance, $\iA((a.c + b.d) \cpar \ol{a}.c) = \{a,b,\ol{a}\}$. Observe that $c$ and $d$, which are not initial, are not included. The initial actions are defined thus: $\iA(\zero_C) = \eset$, $\iA(\alpha \col L.P) = \{\alpha\}$  and $\iA(M + N) = \iA(M) \cup \iA(N)$.

\smallskip 

\noindent \textbf{Antitonicity:} If $C \subseteq C'$ then $\iota(C') \subseteq \iota(C)$. This can be shown be induction on the LTS rules of Fig.~\ref{fig:compatible-rules}. 

\smallskip 

\noindent  \textbf{Clock stability:} The horizon $\clock(P)$ remains stable under derivations, and $P$ can do a $\sigma$ transition only if $\clock(P) = \clockone$. This is a well-formedness condition we impose on process terms.

\smallskip 

\noindent The rules of Fig.~\ref{fig:compatible-rules} use the following auxiliary operations on blocking and predictions.
We combine blocking relations by set union $B_1 \cup B_2$.
The restriction $B \restrict A$ of a blocking relation simply removes the actions $A \cup \ol{A}$ from each entry in $B$, i.e., $B \restrict A \eqdef \{ (C, L {-} (A \cup \ol{A})) \mid (C, L) \in B \}$.
The empty prediction is $\iota^{\eset}$, defined $\iota^{\eset}(C) \eqdef \eset$. As operations on predictions we have point-wise union, written as $\iota_1 + \iota_2$ and defined $(\iota_1+\iota_2)(C) \eqdef \iota_1(C) \cup \iota_2(C)$. We restrict a prediction $\iota$ by removing a set of actions $A \subseteq \A$ pointwise, written $\iota \restrict A$ and defined as $(\iota\restrict A)(C) \eqdef \iota(C) {-} (A \cup \ol{A})$.
With these definitions in place we can take a closer look at the rules in Fig.~\ref{fig:compatible-rules}:
\begin{itemize}
\item (\ConR) This is the standard unfolding rule for constant definitions, including blocking and predictions.
\item (\ComR) The parallel composition rule implements simultaneously the synchronisation of rendezvous actions and the clock: rendezvous synchronisations are dealt with as usual, producing a $\tau$ action. For contrast, broadcast synchronisation, via clocks, produce a clock action (since $\sigma \cpar \sigma = \sigma$), so it remains open for more participants. In either case, the construction of blocking and prediction for the $\ell \cpar \ol{\ell}$ transition is the same: We take the union $B_1 \cup B_2$ of the blockings and the sum $\iota_1 + \iota_2$ of the predictions from the two participating transitions $\ell$ and $\ol{\ell}$.
  In addition we must check the unblocking of the synchronisation $\ell \cpar \ol{\ell}$ cross-wise: the prediction $\iota_i$ of $P_i$ must eschew the blocking $B_{3-i}$. We note that this works uniformly for rendezvous and clock because $\eschews(\iota, B)$ is distributive in both its arguments with respect to union and summation.
\item (\ParR) This rule describes the asynchronous case where one process $P$ in a parallel composition $P \cpar Q$ executes a step unsynchronised with $Q$. The blocking $B$ is inherited from the transition $\alpha$ of $P$. This transition is only enabled as an asynchronous step of $P$ in concurrency with $Q$, if the prediction $\iAstar_{-}(Q)$ eschews the blocking $B$ \textit{and} if the action $\alpha$ is not a clock of $Q$. For if $\alpha = \sigma$ and $ \clock(Q) = \clockone$ then $P$ cannot proceed alone, and it must synchronise lock-step with $Q$ via rule (\ComR). Since the actions predicted by $\iAstar_{-}(Q)$ are in competition to $\alpha$, they are added to the prediction $\iota$.

\item (\ActR) The execution of a prefix $\alpha\col L\cseq P$ publishes the empty prediction function $\iota^{\eset}$, because it does not have any alternative choices or concurrency that would compete. The blocking $\{(\clock(P), L)\}$ lifts the blocking set $L$ of the prefix to a blocking constraint in the scope of the thread's clock. Note, by well-formedness, $\clock(\alpha\col L \cseq P) = \clock(P)$.
\item (\SumR) The iterated application of the summation rule permits us to select any prefix
 $\alpha_n\col L_n \cseq P_n$ of a thread $M = \sum_i \alpha_i\col L_i \cseq P_i$, generating a transition
 $\alpha_n\col B[\iota]$
 of $M$ to $P_n$, with $B = \{(\clock(P_n), L_n)\}$ and $\iota = \iota^\eset \cup \{ \alpha_i \mid i \neq n, \alpha_i \neq \alpha_n \}$. Note that by well-formedness $\clock(P_n) = \clock(M)$. The prediction $\iota$ contains the initial actions $\alpha_i$ for all other choices $i \neq n$ offered by thread $M$ as competitors to $\alpha_n$. Note that $\alpha_n$ does not count as a competitor for itself, whence we subtract it.

 \item (\RestrR) The restriction rule of rendezvous actions $A$ is as in \ccs. It prunes away all local transitions with labels from $A \cup \ol{A}$, because these are no longer available for synchronisation outside. Naturally then, we must remove these labels from the blocking and predictions as well, which is done by the operations $B \restrict A$ and $\iota \restrict A$, defined above.
\end{itemize}
It is immediate to see that Fig.~\ref{fig:compatible-rules} is a conservative extension of Milner's \ccs~\cite{Milner:CCS}.

\subsection{Two Simple Examples}
The interested reader is invited to evaluate the following examples using the LTS. Other examples can be found in \cite{fscd} and with our artefact ({\small\url{https://cstolze.github.io/synpasite/}}).
\begin{itemize}
\item A typical application of priorities is to enforce a ``Read-Before-Write'' policy for a memory cell. The process $S \eqdef w\col w. S + r \col w. S + \sigma \col \{r, w\}. S$ implements the memory cell, with $w$ modelling a write action and $r$ a read, ignoring any data. The process $R \eqdef \ol{r} \cseq 0_\clockone$ implements a reader and $W \eqdef \ol{w} \cseq 0_\clockone$ a writer. The composition $R \cpar W \cpar S$ will first do a write, then a read. The memory with two writers $W \cpar W \cpar S$ will block and with two readers $R \cpar R \cpar S$ will permit the readers to move in any order.
\item Clocks add expressivity when used with priorities. Take for instance the processes $\mathsf{p} \eqdef r \col \ol{w} \cseq \mathsf{p}$ and $\mathsf{q} \eqdef w \cseq \mathsf{q}$. The process $\mathsf{p} \cpar \mathsf{q}$ will never do $r$, because $w$ is always in the prediction horizon. However, if you consider $\mathsf{p}' \eqdef r \col \ol{w} \cseq \sigma \cseq \mathsf{p}'$ and $\mathsf{q}' \eqdef w \cseq \sigma \cseq \mathsf{q}'$, then the process $\mathsf{p}' \cpar \mathsf{q}'$ will first do $w$, then $r$, then $\sigma$, and then loop from start  $\mathsf{p} \cpar \mathsf{q}$. In this fashion we can model iterated computations on shared memory.
\end{itemize}



\section{From Confluence to Coherence}
\label{sec:coherence}

\begin{figure}
\begin{center}
  \includegraphics[scale=0.7]{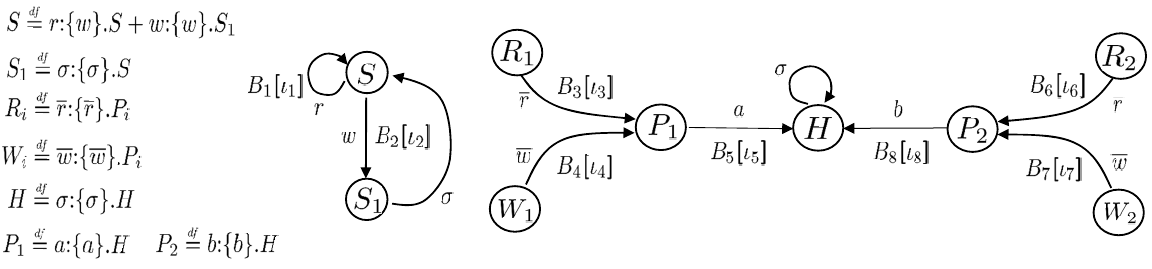}
\end{center}
\caption{Store $S$, Readers $R_i$ and Writers $W_i$, in \ccslm  (left) and strategic transitions (right).}
\label{fig:store-readers}
\end{figure}

We now finally come to lift and generalise Milner's classical notion of confluence (Def.~\ref{def:local-confluence}) for stategic transitions. For motivation we will refer to the examples in Fig.~\ref{fig:store-readers}.

Classical determinacy (Def.~\ref{def:determinacy}) requires that any two transitions with the same visible action must lead to congruent states, which is too strong: The composition of two readers $R_1 \cpar R_2 = \ol{r} \cseq P_1 \cpar \ol{r} \cseq P_2 = \ol{r} \cseq a \cseq \zero \cpar \ol{r} \cseq b \cseq \zero$, sharing the same memory access point generates $\ol{r}$-labelled transitions to non-congruent successor states $P_1 \cpar R_2 \not\cong R_1 \cpar P_2$, in general.
Using strategic labels we can disambiguate them. The strategic transitions
$$
  R_1 \cpar R_2 \Derives{\ol{r}}{\iota_3'}{B_3} P_1 \cpar R_2
\text{ and }
  R_1 \cpar R_2 \Derives{\ol{r}}{\iota_6'}{B_6} R_1 \cpar P_2
\text{ with }
  \iota_3' = \iota_3 + \iAstar_{-}(R_2) = \iAstar_{-}(R_2)
\text{ and }
  \iota_6' = \iota_6 + \iAstar_{-}(R_1) = \iAstar_{-}(R_1)
$$
are obtained by the (\ParR) rule and have $\ol{r}$ in both predictions since $\ol{r} \in \iAstar_{\clockone}(R_2)$ and $\ol{r} \in \iAstar_{\clockone}(R_1)$.
So, in each transition, the other label $\ol{r}$ is observable as a competitor within the clock horizon, i.e., $\ol{r} \in \iota_3'(\clockone)$ and $\ol{r} \in \iota_6'(\clockone)$.
With this in mind, we replace `determinate' (Def.~\ref{def:determinacy}) by the following notion of `observable'.

\begin{definition}
  A process $P$ is \textit{observable} (modulo $\cong$) if for all derivatives $Q$ and transitions with $Q \Derives{\alpha_1}{\iota_1}{B_1} Q_1$ and $Q \Derives{\alpha_2}{\iota_2}{B_2} Q_2$,
  such that $\alpha_1 \neq \alpha_2$ or $Q_1 \not\cong Q_2$, if $\{\alpha_1, \alpha_2\} \subseteq \L$ then $\alpha_1 \in \iota_{2}(\clockone)$ and $\alpha_2 \in \iota_{1}(\clockone)$.
\label{def:observable}
\end{definition}
\noindent
Note that Def.~\ref{def:observable} permits silent transitions $\alpha_1 = \tau = \alpha_2$ to end up in non-congruent states $Q_1 \not\cong Q_2$. This is needed for strong bisimulation. For instance, $R_1 \cpar S \cpar R_2$ has $\tau$-transitions to non-bisimilar states $P_1 \cpar S \cpar R_2  \not\sim R_1 \cpar S \cpar P_2$. This is non-determinate by Def.~\ref{def:local-confluence}, but observable by Def.~\ref{def:observable}.
Also, since a clock $\alpha_1 = \sigma = \alpha_2$ is never in competition with itself, i.e., $\sigma \not \in \iota_{i}(\clockone)$, then observability forces $Q_1 \cong Q_2$, i.e., the clock is deterministic. In this way, we capture time determinacy of timed extensions of \ccs~\cite{TPL,ATP}.

The classical notion of confluence requires reconvergence for \textit{any pair} of distinctly labelled transitions. This is too strong for our purposes.
A shared store $S$, by its very nature, must offer a preempting choice for a reader (on $r$) and a writer (on $w$), generating a race that leads to incongruent successor states $S \not\cong S_1$. The read and the write are distincly labelled but not confluent by Def.~\ref{def:local-confluence}. But this is no problem, if the store resolves the race by priority, giving $w$ precedence over $r$ (or the other way around). Technically, this is done by adding the blocking constraint $(\clockone, \{w\}) \in B_1$ in the stategic action $r \col B_1[\iota_1]$ of the read (see Fig.~\ref{fig:store-readers}).
The presence of a concurrent writer $W_1$, say, with $\ol{w} \in \iAstar_\clockone(W_1)$ in its prediction, will then block the read action $r \col B_1[\iota_1]$ in rule (\ParR) since $\eschews(\iAstar_{-}(W_1),B_1)$ is false. Hence, there is no competition between the two distinct choices $r$ and $w$ of $S$ and we do not need to require reconvergence, unlike with Def.~\ref{def:local-confluence}.

The classical notion of confluence is also too weak, when it comes to sharing, as it requires reconvergence for any pair of \textit{distinctly labelled} transitions. The write label $w$ of $S$ can be consumed only once, by a single writer. Two writers $S \cpar W_1 \cpar W_2$ with $W_1 = \ol{w} \cseq P_1$ and $W_2 = \ol{w} \cseq P_2$ generate a race condition.
Unfortunately, the standard test for confluence will not detect the problem, because the $w$-transition of $S$ is not distinct from itself. But competes against itself, because it can be consumed only once. Here, we eliminate the race by making the strategic action $w \col B_2[\iota_2]$ be self-blocking, so that $w \in [B_2](\clockone)$. This blocks the write action in the presence of a second writer. With a single writer, $S \cpar W_1$ reduces to $S_1 \cpar P_1$ but $S \cpar W_1 \cpar W_2$ blocks in (\ParR). For the reading action $r$ we have a different situation. The store happily supports multiple readers, as $r$ may be infinitely repeated without $S$ changing its state. In this case, the stategic action $r \col B_1[\iota_1]$ can be such that $r \not\in [B](\clockone)$.

In sum, we must require reconvergence not just for any pair of distinctly  labelled transitions but any pair of mutually non-blocking transitions.
The following notion of `independence' replaces the `distinctness' condition in the classical notion of confluence, taking into account the blocking relation.

\begin{definition}[Independence]
  Two transitions $Q \Derives{\alpha_1}{\iota_1}{B_1} Q_1$ \text{ and } $Q \Derives{\alpha_2}{\iota_2}{B_2} Q_2$ are \textit{independent} (modulo $\cong$) if $\{\alpha_1, \alpha_2\} = \{\sigma\}$ or $\{\alpha_1, \alpha_2\} \subseteq \R \cup \{\tau\}$ and
  one of the following holds:
  \begin{enumerate}
  \item $\{\alpha_1, \alpha_2\} \neq \{\tau\}$ and both $\alpha_1 \not\in [B_2](\clockone)$ and $\alpha_2 \not\in [B_1](\clockone)$, or
  \item $\alpha_1 = \alpha_2$ and $Q_1 \not\cong Q_2$.
 \end{enumerate}
\label{def:independence-new-old}
\end{definition}
\noindent We now define \textit{coherence} by refactoring the classical notion of confluence (Def.~\ref{def:local-confluence}) with `observable' for `determinate' and `independent' for `distinct'.
In addition, we require that the reconvergence is monotonic in the blocking relations and prediction functions.
\begin{definition}[Coherence] \mbox{}
  A process $Q$ satisfies the Diamond Property (\CR) if, for every pair of independent transitions, 
  $$
    Q \Derives{\alpha_1}{\iota_1}{B_1} Q_1
    \text{ and }
    Q \Derives{\alpha_2}{\iota_2}{B_2} Q_2,
  $$
  called a \textit{divergence} (modulo $\cong$),
  there exists a \textit{reconvergence} (modulo $\cong$), consisting of processes $Q_1'$, $Q_2'$ with $Q_1' \cong Q_2'$ and
  $$
    Q_1 \Derives{\alpha_2}{\iota_2'}{B_2'} Q_1'
    \text{ and }
    Q_2 \Derives{\alpha_1}{\iota_1'}{B_1'} Q_2'
  $$
  such that $B_i' \sqsubseteq B_i$ and if $\alpha_i \in \R$ then $\iota_i' \sqsubseteq \iota_i$ for $i \in \{ 1, 2 \}$.
  $P$ is \textit{coherent} (modulo $\cong$) if it is observable (modulo $\cong$) and every derivative of $P$ satisfies \CR.
 \label{def:coherence}
\end{definition}
\noindent The most important consequence of coherence is that coherent process are ``determinate'' under silent actions, i.e., reduce to a unique normal form. As in the $\lambda$-calculus, we call a process $N$ in \textit{normal form} if it does not have any $\tau$-transitions. Then, it is internally stable without rendezvous synchronisations, yet may still participate in a clock step. Let us write $P \Downarrow N$ to express that $P$ reduces to normal form $N$.
\begin{lemma}
  Let $P$ be coherent. Then $P \Downarrow N_1$ and $P \Downarrow N_2$ implies $N_1 \cong N_2$.
\label{thm:church-rosser}
\end{lemma}
\noindent Lem.~\ref{thm:church-rosser} only concerns the reductions, i.e., the $\tau$-transitions, of a coherent process. For races between $\tau$ and $\sigma$, coherence does not imply reconvergence, for good reasons. We can thus model scenarios where a synchronous system (horizon $\clockone$) communicates asynchronously with a process of horizon $\clockzero$. Such interactions in general create data races for good reason: A synchronous system can notice if a communication with external processes happens before or after the clock, and act differently.
Our main result is the following Preservation Theorem, stating that parallel composition and restriction preserve coherence.
\begin{theorem}[Preservation] If $P_1$, $P_2$ are coherent then $P_1 \cpar P_2$ is coherent. If $P$ is coherent, then $P \restrict A$ is coherent.
\label{thm:cpar-restr-coherent}
\end{theorem}
\noindent How to we establish coherence? To prove that a given process $P \in \cP$ is coherent requires that we show \CR for all derivatives of $P$. The general technique to prove membership then is co-induction on the immediate transitions of a process. Formally, we call a class of processes $\mathsf{Coh} \subseteq \cP$ a \textit{coherence class} if it is (i) derivation closed, i.e., if $Q \in \mathsf{Coh}$ and $Q \fstep{} Q'$ implies $Q' \in \mathsf{Coh}$, (ii) all $Q \in \mathsf{Coh}$ are observable and satisfy \CR. It is easy to see that $P$ is coherent iff there exists a coherence class $\mathsf{Coh}$ with $P \in \mathsf{Coh}$. The standard application of this principle is to take $\mathsf{Coh}$ to be the set of all derivatives of $P$, which is trivially derivation closed, and show that each of them is observable and satisfies \CR.

We illustrate this technique on our store-reader-writer example which form the transition system seen in Fig.~\ref{fig:store-readers} with the transitions decorated with their strategic labels of shape $\alpha \col B[\iota]$.
All processes are synchronous with $\clock(S) = \clock(W_i) = \clock(R_i) = \clockone$, so each blocking constraint $(C, L) \in B$ has $C = \clockone$, whence $[B](\clockzero) = \eset$. For notational conciseness we identify $B$ with the set $[B](\clockone) \subseteq \L$, i.e., the set of actions that block $\alpha$, within the current clock cycle.
Likewise, we can identify the prediction $\iota$ with the set $\iota(\clockone) \subseteq \L$, since $\iota(\clockzero) = \iota(\clockone)$. The set $\iota(\clockzero)$ is not larger than $\iota(\clockone)$ since there are no asynchronous actions, i.e., with clock horizon $\clockzero$.
The fact that our basic processes model single threads (no internal concurrency) means that the $\iota$ of an action is essentially the set of distinct actions in immediate competition with $\alpha$ in the respective thread, generated by rules (\SumR) and (\ActR).
For instance the initial looping transition $r\col B_1[\iota_1]$ of $S$ in Fig.~\ref{fig:store-readers} has $B_1 = \{ w \}$ because we have write-before-read and $\iota_1 = \{ w\}$ because $S$ has an initial $w$-transition competing with $r$.
The strategic label $r\col B_1[\iota_1]$ is thus $r\col \{w\}[\{w\}]$ generated from the syntactic prefix $r \col \{ w\} \cseq S$ of the thread $S = r \col \{w\} \cseq S + w \col \{w\} \cseq S_1$.

Now let us check coherence.
Firstly, both derivatives $S$ and $S_1$ are observable. In fact, the only two transitions to consider with $\alpha_1 \neq \alpha_2$ or $Q_1 \not\cong Q_2$ are $\alpha_1 = r$ and $\alpha_2 = w$ out of $S$. But they satisfy $\alpha_1 \in \iota_2 = \{r\}$ and $\alpha_2 \in \iota_1 = \{w\}$. Next we check that $S$ satisfies \CR.
Since the $r$ is blocked by $w$ and the $w$-transition is self-blocking, the only pair of independent transitions out of $S$ is the $r$-transition competing against itself, $S \Derives{\alpha_1}{\iota_1}{B_1} Q_1$ and $S \Derives{\alpha_2}{\iota_2}{B_2} Q_2$ with $Q_1 = S = Q_2$,
$\alpha_1 = r = \alpha_2$ and $B_1 = \{w\} = \iota_1$ and $B_2 = \{w\} = \iota_2$.
But the $r$-loop can be infinitely repeated and so both have a trivial reconvergence with
$Q_1 \Derives{\alpha_2}{\iota_2'}{B_2'} S$ and $Q_2 \Derives{\alpha_1}{\iota_1'}{B_1'} S$ where $B_2' = \{w\} = \iota_2'$ and $B_1' = \{w\} = \iota_1'$  which trivially satisfies $S \cong S$, $B_i' \sqsubseteq B_i$ and $\iota_i' \sqsubseteq \iota_i'$ as required by Def.~\ref{def:coherence}. Finally note that $S_1$ has a single $\sigma$-transition which is self-blocking and thus trivially satisfies \CR. In the same fashion, one verifies that all processes and derivatives of Fig.~\ref{fig:store-readers} are observable and satisfy \CR.

Observe that the priority system allows us to check for absence: if $\alpha \col L\cseq P$
can proceed, we are sure that none of the actions in $L$
can synchronise up to $\clock(P)$. Using $\tau$ prefixes allows us to check for absence without doing anything else: For instance, $a \col \{\} \cseq P + \tau \col\{a\} \cseq Q$ intuitively means ``if $a$ is feasible in the current cycle, then do $a$ then $P$, else do $Q$''.



\section{Conclusion} \label{sec:end}

In this paper we propose the novel concept of coherence to strengthen the notion of confluence from Milner's classical  theory~\cite{Milner:CCS}. It is based on the mechanism of strategic action labels that generalises earlier work on priority-guarded \ccs~\cite{CamilleriWin95,CleavelandLN01,Phillips08}. Our Preservation Theorem~\ref{thm:cpar-restr-coherent} is a significant advance. It holds without additional conditions unlike Milner's notion of confluence, which needs sort separation. In addition, it applies to $P \cpar Q$ and $P \restrict A$ separately, unlike confluence that only holds for a combination of both, and it applies to any congruence $\cong$. This extends the confluence theory of Chap.~11 of Milner's book. By adding clocks we can treat determinacy compositionally in multi-threaded shared memory and synchronous programming with reaction to absence à la Esterel. Previous attempts such as the S$\pi$ language~\cite{Amadio:IC07}, or SPL~\cite{LuttgenBC99} do not adequatly encode Esterel constructs. For example, S$\pi$ does not deal with immediate reaction to absence and SPL only encodes local consistency rather than Esterel's global consistency.

It is known from priority-guarded extensions of \ccs that strategic action labels add expressiveness~\cite{Versari-etal:MSCS09}.
In the case of \ccslm it is easy to see that there is no compositional encoding into \ccs. Consider the terms $\ol{a}\col\ol{a}$ and $a\col a$ which have one transition each, with action $\ol{a}$ and $a$, respectively. When composed in parallel, $\ol{a}\col\ol{a} \cpar a\col a$ has only one possible transition, performing $\tau$. However, there are no \ccs processes $P$ and $Q$ which perform one transition each, and, put in parallel, only perform one transition, because parallel composition in \ccs preserves the transitions of the composed processes.
What is noteworthy is that unlike in traditional priority-guarded \ccs such as \ccsp, there does not seem to be a natural expansion lemma for the stategic labels of \ccslm under the scheduling rules of Fig.~\ref{fig:compatible-rules}, i.e., we cannot hope to rewrite every parallel composition as a single thread. The reason is that a self-blocking prefix $a \col L \cseq P$ with $a \in L$ will block when put in parallel with two consumer threads $M_1$ and $M_2$ with $\ol{a} \in \iAstar(M_i)$, but it does not block with only one of them. So, if the composition $M_1 \cpar M_2$ was expandable to a congruent single-threaded process $M$, then $(a \col L) \cpar M_1 \cpar M_2 \cong (a \col L) \cpar M$ and so $(a \col L) \cpar M_1 \cpar M_2$ should not block.
Interestingly, self-blocking prefixes have not been considered in the classical theories~\cite{CamilleriWin95,CleavelandLN01,Phillips08}, while here they naturally play an important role. They bring about a true-concurrency semantics.

Moreover, in these classical approaches,  prediction is based on the immediate actions $\iA$ of a process, with the effect that the scheduling priorities can change with each transition.
Here, we use $\iAstar$ in the scheduling rules which is more conservative (Constructivity Assumption) and enjoys a useful monotonicity property: once an action is unblocked, it remains unblocked until the clock horizon is reached.
This has the interesting consequence that summation $+$ does not seem to be expressible anymore, in terms of prioritised prefixes and restricted parallel composition, unlike in \ccsp.
To be specific, define $a \cseq P \oplus b \cseq Q$ as an abbreviation of $(a \col c \cseq (P \cpar  \ol{c})) \cpar (b \col c \cseq (Q \cpar \ol{c})) \restrict c$. Then, in  \ccsp this sum $\oplus$ indeed acts like a non-deterministic free choice $a \cseq P + b \cseq Q$~\cite{Phillips08}.
In our semantics, however, $\oplus$ behaves deterministically. It blocks, because of the conservative ``up-to-clocks'' predictions $\{a, \ol{c}\} \subseteq \iAstar(a \col c \cseq (P \cpar \ol{c}))$ which can see the blocking action $\ol{c}$ even though it is guarded by the $a$ action.
Moving away from \textit{immediate enabling} based on $\iA$ to \textit{up-to-clocks enabling} based on $\iAstar$ is the cornerstone for modelling the synchronous micro-macro step abstraction of the constructive semantics of Esterel-style languages~\cite{Berry99,vonHanxledenDM+14,BoussinotD96}.
Let us note that the classical immediate enabling $\iA$ can be used to code synchronous Statecharts as has been shown by~\cite{LuttgenBC99} using a \ccs-style algebra.
But this only captures a weak form of reaction to absence (with local consistency rather than global consistency) and depends on special syntactic operators.

We leave the exact characterisation of expressiveness of \ccslm as an open problem. We also plan to expand our theory by standard instruments such as notions of bisimulation, observational congruence, and associated algebraic axiomatisations. We are also working on extending the theory to other static operators such as clock hiding (for time abstraction) and to multiple clocks which will apply to globally-asynchronous, locally-synchronous programming and extend our earlier work on \pmc~\cite{AndersenMen94} and \csa~\cite{CleavelandLM97}.
In \pmc there are no priorities and the timeout behaves simply like a choice $P + \sigma\col\eset \cseq Q$ in \ccslm. For contrast, \csa uses a fixed priority scheme that makes rendezvous actions take priority over any clock. In \ccslm the timeout can be coded, too, as a sum $P + \sigma\col \L \cseq Q$. Thus, by permitting general priority schemes as in \ccsp, which has no clocks, \ccslm is likely more expressive than any of these prior systems.

We point out that the general principle of LTS with negative premises has been studied in~\cite{BolGroote:ACM96}.
The priority guards here and in Phillips' work are special, however, because they are not negations of the transition relation itself but they are defined independently via predicates such as $\iA$ and $\iAstar$. This avoids many a complication of the general theory.
Priorities as negative premises for system specification are also fundamental in the \bip algebra of interactions proposed in~\cite{BliudzeSifakis2007} which present a generic and compositional paradigm of coordination protocols for parallel languages.
The synchronisation mechanisms of \ccs rendezvous and \csp broadcast actions (clocks) are but special cases of interactions on \bip connectors.
\bip interactions (defined on distinct alphabets of components) can distinguish the identity of threads. Hence, \bip can capture the semantics of self-blocking as in \ccslm, where $a \col a \cpar \ol{a}$ is not blocking while $a \col a \cpar \ol{a} \cpar \ol{a}$ is blocking. This is not possible in \ccsp, which does not have self-blocking at all. However, in \bip we would still only capture \ccsp-style local priorities, based on immediate initial actions, not priorities ``up-to-clocks'' like in \ccslm.
An even more important open question on \bip expressiveness stems from the nature of
\bip priorities, which require that an interaction has to be \textit{maximal} in some strict ``priority'' ordering. It will be interesting to investigate if priorities in the style of \ccslm (and a fortiori of prioritised \ccs) can actually be expressed in this way.  We are not aware of any comparison of \bip~\cite{BliudzeSifakis2007} with prioritised CCS~\cite{CamilleriWin95,CleavelandLM97,CLN98,Phillips08}. Reciprocally, we leave it to future work to explore how our notion of coherence carry over to general coordination languages such as \bip or \reo~\cite{Dokter-etal:JLAP17}.


\bigskip 

\noindent
\textbf{Acknowledgements.}
The authors are grateful to Martin Steffen, Gerald Lüttgen, Robert de Simone, Adrien Guatto, and Furio Honsell for useful discussions and comments. Michael Mendler and Claude Stolze are funded by DFG under grant number ME 1427/7-1, Luigi Liquori is funded by ETSI under grant number STF 655.

\bibliography{references-places}

\begin{thebibliography}{10}
\providecommand{\bibitemdeclare}[2]{}
\providecommand{\surnamestart}{}
\providecommand{\surnameend}{}
\providecommand{\urlprefix}{Available at }
\providecommand{\url}[1]{\texttt{#1}}
\providecommand{\href}[2]{\texttt{#2}}
\providecommand{\urlalt}[2]{\href{#1}{#2}}
\providecommand{\doi}[1]{doi:\urlalt{https://doi.org/#1}{#1}}
\providecommand{\eprint}[1]{arXiv:\urlalt{https://arxiv.org/abs/#1}{#1}}
\providecommand{\bibinfo}[2]{#2}

\bibitemdeclare{article}{Amadio:IC07}
\bibitem{Amadio:IC07}
\bibinfo{author}{R.~M. \surnamestart Amadio\surnameend} (\bibinfo{year}{2007}):
  \emph{\bibinfo{title}{A Synchronous $\pi$-Calculus}}.
\newblock {\slshape \bibinfo{journal}{Information and Computation}}
  \bibinfo{volume}{9}(\bibinfo{number}{205}), pp. \bibinfo{pages}{1470--1490},
  \doi{10.1016/j.ic.2007.02.002}.

\bibitemdeclare{inproceedings}{AndersenMen94}
\bibitem{AndersenMen94}
\bibinfo{author}{H.~R. \surnamestart Andersen\surnameend} \&
  \bibinfo{author}{M.~\surnamestart Mendler\surnameend} (\bibinfo{year}{1994}):
  \emph{\bibinfo{title}{An asynchronous process algebra with multiple clocks}}.
\newblock In: {\slshape \bibinfo{booktitle}{Proc.~{ESOP}}}, pp.
  \bibinfo{pages}{58--73}, \doi{10.1007/3-540-57880-3\_4}.

\bibitemdeclare{book}{Berry99}
\bibitem{Berry99}
\bibinfo{author}{G.~\surnamestart Berry\surnameend} (\bibinfo{year}{1999}):
  \emph{\bibinfo{title}{The Constructive Semantics of Pure {Esterel}}}.
\newblock \bibinfo{publisher}{Draft Book}.

\bibitemdeclare{inproceedings}{BliudzeSifakis2007}
\bibitem{BliudzeSifakis2007}
\bibinfo{author}{S.~\surnamestart Bliudze\surnameend} \&
  \bibinfo{author}{J.~\surnamestart Sifakis\surnameend} (\bibinfo{year}{2007}):
  \emph{\bibinfo{title}{The algebra of connectors: structuring interaction in
  {BIP}}}.
\newblock In: {\slshape \bibinfo{booktitle}{Proc. of EMSOFT}},
  \bibinfo{publisher}{{ACM}}, pp. \bibinfo{pages}{11--20},
  \doi{10.1145/1289927.1289935}.

\bibitemdeclare{article}{BolGroote:ACM96}
\bibitem{BolGroote:ACM96}
\bibinfo{author}{R.~\surnamestart Bol\surnameend} \& \bibinfo{author}{J.~F.
  \surnamestart Groote\surnameend} (\bibinfo{year}{1996}):
  \emph{\bibinfo{title}{The Meaning of Negative Premises in Transitions System
  Specifications}}.
\newblock {\slshape \bibinfo{journal}{{JACM}}}
  \bibinfo{volume}{43}(\bibinfo{number}{5}), pp. \bibinfo{pages}{863--914},
  \doi{10.1145/234752.234756}.

\bibitemdeclare{article}{BoussinotD96}
\bibitem{BoussinotD96}
\bibinfo{author}{F.~\surnamestart Boussinot\surnameend} \&
  \bibinfo{author}{R.~\surnamestart de~Simone\surnameend}
  (\bibinfo{year}{1996}): \emph{\bibinfo{title}{The {SL} synchronous
  language}}.
\newblock {\slshape \bibinfo{journal}{IEEE Transactions on Software
  Engineering}} \bibinfo{volume}{22}(\bibinfo{number}{4}), pp.
  \bibinfo{pages}{256--266}, \doi{10.1109/32.491649}.

\bibitemdeclare{article}{CamilleriWin95}
\bibitem{CamilleriWin95}
\bibinfo{author}{J.~\surnamestart Camilleri\surnameend} \&
  \bibinfo{author}{G.~\surnamestart Winskel\surnameend} (\bibinfo{year}{1995}):
  \emph{\bibinfo{title}{{CCS} with priority choice}}.
\newblock {\slshape \bibinfo{journal}{Information and Computation}}
  \bibinfo{volume}{116}(\bibinfo{number}{1}), pp. \bibinfo{pages}{26--37},
  \doi{10.1006/INCO.1995.1003}.

\bibitemdeclare{inproceedings}{CleavelandLM97}
\bibitem{CleavelandLM97}
\bibinfo{author}{R.~\surnamestart Cleaveland\surnameend},
  \bibinfo{author}{G.~\surnamestart L{\"u}ttgen\surnameend} \&
  \bibinfo{author}{M.~\surnamestart Mendler\surnameend} (\bibinfo{year}{1997}):
  \emph{\bibinfo{title}{An Algebraic Theory of Multiple Clocks}}.
\newblock In: {\slshape \bibinfo{booktitle}{Proc. of {CONCUR}}}, {\slshape
  \bibinfo{series}{LNCS}} \bibinfo{volume}{1243}, pp.
  \bibinfo{pages}{166--180}, \doi{10.1007/3-540-63141-0\_12}.

\bibitemdeclare{article}{CLN98}
\bibitem{CLN98}
\bibinfo{author}{R.~\surnamestart Cleaveland\surnameend},
  \bibinfo{author}{G.~\surnamestart L{\"{u}}ttgen\surnameend} \&
  \bibinfo{author}{V.~\surnamestart Natarajan\surnameend}
  (\bibinfo{year}{1998}): \emph{\bibinfo{title}{A Process Algebra with
  Distributed Priorities}}.
\newblock {\slshape \bibinfo{journal}{Theor. Comput. Sci.}}
  \bibinfo{volume}{195}(\bibinfo{number}{2}), pp. \bibinfo{pages}{227--258},
  \doi{10.1016/S0304-3975(97)00221-1}.

\bibitemdeclare{article}{Dokter-etal:JLAP17}
\bibitem{Dokter-etal:JLAP17}
\bibinfo{author}{K.~\surnamestart Dokter\surnameend}, \bibinfo{author}{S.-S.
  \surnamestart Jongmans\surnameend}, \bibinfo{author}{F.~\surnamestart
  Arbab\surnameend} \& \bibinfo{author}{S.~\surnamestart Bliudze\surnameend}
  (\bibinfo{year}{2017}): \emph{\bibinfo{title}{Combine and Conquer: Relating
  {BIP} and {Reo}}}.
\newblock {\slshape \bibinfo{journal}{Journal of Logical and Algebraic Methods
  in Programming}} \bibinfo{volume}{86}, pp. \bibinfo{pages}{134--156},
  \doi{10.1016/j.jlamp.2016.09.008}.

\bibitemdeclare{inproceedings}{vonHanxledenDM+14}
\bibitem{vonHanxledenDM+14}
\bibinfo{author}{R.~\surnamestart von Hanxleden\surnameend},
  \bibinfo{author}{B.~\surnamestart Duderstadt\surnameend},
  \bibinfo{author}{C.~\surnamestart Motika\surnameend},
  \bibinfo{author}{S.~\surnamestart Smyth\surnameend},
  \bibinfo{author}{M.~\surnamestart Mendler\surnameend},
  \bibinfo{author}{J.~\surnamestart Aguado\surnameend},
  \bibinfo{author}{S.~\surnamestart Mercer\surnameend} \&
  \bibinfo{author}{O.~\surnamestart O'Brien\surnameend} (\bibinfo{year}{2014}):
  \emph{\bibinfo{title}{{SCCharts: Sequentially Constructive Statecharts} for
  Safety-Critical Applications}}.
\newblock In: {\slshape \bibinfo{booktitle}{Proc. of PLDI}},
  \bibinfo{publisher}{ACM}, pp. \bibinfo{pages}{372--383},
  \doi{10.1145/2594291.2594310}.

\bibitemdeclare{article}{TPL}
\bibitem{TPL}
\bibinfo{author}{M.~\surnamestart Hennessy\surnameend} \&
  \bibinfo{author}{T.~\surnamestart Regan\surnameend} (\bibinfo{year}{1995}):
  \emph{\bibinfo{title}{A process algebra for timed system}}.
\newblock {\slshape \bibinfo{journal}{Information and Computation}}
  \bibinfo{volume}{117}, pp. \bibinfo{pages}{221--239},
  \doi{10.1006/INCO.1995.1041}.

\bibitemdeclare{book}{Hoare:CSP}
\bibitem{Hoare:CSP}
\bibinfo{author}{C.A.R. \surnamestart Hoare\surnameend} (\bibinfo{year}{1985}):
  \emph{\bibinfo{title}{Communicating Sequential Processes}}.
\newblock \bibinfo{publisher}{Prentice-Hall}.
\newblock \bibinfo{note}{256 pages}.

\bibitemdeclare{misc}{fscd}
\bibitem{fscd}
\bibinfo{author}{L.~\surnamestart Liquori\surnameend},
  \bibinfo{author}{M.~\surnamestart Mendler\surnameend} \&
  \bibinfo{author}{C.~\surnamestart Stolze\surnameend} (\bibinfo{year}{2026}):
  \emph{\bibinfo{title}{A process calculus with clocks and priorities}}.
\newblock \bibinfo{howpublished}{\url{https://inria.hal.science/hal-05497982}}.

\bibitemdeclare{inproceedings}{LuttgenBC99}
\bibitem{LuttgenBC99}
\bibinfo{author}{G.~\surnamestart L{\"{u}}ttgen\surnameend},
  \bibinfo{author}{M.~\surnamestart von~der Beeck\surnameend} \&
  \bibinfo{author}{R.~\surnamestart Cleaveland\surnameend}
  (\bibinfo{year}{1999}): \emph{\bibinfo{title}{Statecharts Via Process
  Algebra}}.
\newblock In: {\slshape \bibinfo{booktitle}{Proc. of {CONCUR}}}, {\slshape
  \bibinfo{series}{LNCS}} \bibinfo{volume}{1664},
  \bibinfo{publisher}{Springer}, pp. \bibinfo{pages}{399--414},
  \doi{10.1007/3-540-48320-9\_28}.

\bibitemdeclare{article}{SCCS}
\bibitem{SCCS}
\bibinfo{author}{R.~\surnamestart Milner\surnameend} (\bibinfo{year}{1983}):
  \emph{\bibinfo{title}{Calculi for Synchrony and Asynchrony}}.
\newblock {\slshape \bibinfo{journal}{Theor. Comput. Sci.}}
  \bibinfo{volume}{25}, pp. \bibinfo{pages}{267--310},
  \doi{10.1016/0304-3975(83)90114-7}.

\bibitemdeclare{book}{Milner:CCS}
\bibitem{Milner:CCS}
\bibinfo{author}{R.~\surnamestart Milner\surnameend} (\bibinfo{year}{1989}):
  \emph{\bibinfo{title}{Communication and Concurrency}}.
\newblock \bibinfo{publisher}{Prentice Hall}.

\bibitemdeclare{article}{ATP}
\bibitem{ATP}
\bibinfo{author}{X.~\surnamestart Nicollin\surnameend} \&
  \bibinfo{author}{J.~\surnamestart Sifakis\surnameend} (\bibinfo{year}{1994}):
  \emph{\bibinfo{title}{The Algebra of Timed Processes, {ATP:} Theory and
  Application}}.
\newblock {\slshape \bibinfo{journal}{Inf. Comput.}}
  \bibinfo{volume}{114}(\bibinfo{number}{1}), pp. \bibinfo{pages}{131--178},
  \doi{10.1006/INCO.1994.1083}.

\bibitemdeclare{article}{Phillips08}
\bibitem{Phillips08}
\bibinfo{author}{I.~\surnamestart Phillips\surnameend} (\bibinfo{year}{2008}):
  \emph{\bibinfo{title}{CCS with priority guards}}.
\newblock {\slshape \bibinfo{journal}{The Journal of Logic and Algebraic
  Programming}} \bibinfo{volume}{75}(\bibinfo{number}{1}), pp.
  \bibinfo{pages}{139--165}, \doi{10.1016/J.JLAP.2007.06.005}.

\bibitemdeclare{incollection}{CleavelandLN01}
\bibitem{CleavelandLN01}
\bibinfo{author}{G.~L\"uttgen \surnamestart R.~Cleaveland\surnameend} \&
  \bibinfo{author}{V.~\surnamestart Natarajan\surnameend}
  (\bibinfo{year}{2001}): \emph{\bibinfo{title}{Priority in Process Algebra}}.
\newblock In: {\slshape \bibinfo{booktitle}{Handbook of Process Algebra}},
  \bibinfo{publisher}{North-Holland / Elsevier}, pp. \bibinfo{pages}{711--765},
  \doi{10.1016/B978-044482830-9/50030-8}.

\bibitemdeclare{article}{Versari-etal:MSCS09}
\bibitem{Versari-etal:MSCS09}
\bibinfo{author}{C.~\surnamestart Versari\surnameend},
  \bibinfo{author}{M.~\surnamestart Busi\surnameend} \&
  \bibinfo{author}{R.~\surnamestart Gorrieri\surnameend}
  (\bibinfo{year}{2009}): \emph{\bibinfo{title}{An expressiveness study of
  priority in process calculi}}.
\newblock {\slshape \bibinfo{journal}{Mathematical Structures in Computer
  Science}} \bibinfo{volume}{6}(\bibinfo{number}{19}), pp.
  \bibinfo{pages}{1161–--1189}, \doi{10.1017/S0960129509990168}.

\end{thebibliography}
\end{document}